\documentclass[12pt,a4paper,final]{iopart}

\usepackage{iopams}  
\usepackage{graphicx}
\usepackage[breaklinks=true,colorlinks=true,linkcolor=blue,urlcolor=blue,citecolor=blue]{hyperref}
\usepackage{booktabs}
\expandafter\let\csname equation*\endcsname\relax
\expandafter\let\csname endequation*\endcsname\relax
\usepackage{amsmath}
\usepackage[]{algorithm2e}
\PassOptionsToPackage{linktocpage}{hyperref}

\newcommand{\defeq}{:=}
\newcommand{\TA}{\tau_A}
\newcommand{\TB}{\tau_B}

\begin{document}

\title[Co-Diffusion of Social Contagions]{Co-Diffusion of Social Contagions}

\author{Ho-Chun Herbert Chang}
\address{Department of Mathematics, Dartmouth College}
\ead{herbert.18@dartmouth.edu}

\author{Feng Fu}
\address{Department of Mathematics, Dartmouth College}
\ead{feng.fu@dartmouth.edu}

\begin{abstract}
Prior social contagion models consider the spread of either one contagion at a time on interdependent networks or multiple contagions on single layer networks or under assumptions of pure competition. We propose a new threshold model for the diffusion of multiple contagions. Individuals are placed on a multiplex network with a periodic lattice and a random-regular-graph layer. On these population structures, we study the interface between two key aspects of the diffusion process: the level of synergy between two contagions, and the rate at which individuals become dormant after adoption. Dormancy is defined as a looser form of immunity that models the ability to spread without resistance. 
Monte Carlo simulations reveal lower synergy makes contagions more susceptible to percolation, especially those that diffuse on lattices. Faster diffusion of one contagion with dormancy probabilistically blocks the diffusion of the other, in a way similar to ring vaccination. We show that within a band of synergy, contagions on the lattices undergo bimodal or trimodal branching if they are the slower diffusing contagion.

\end{abstract}

\pacs{00.00, 20.00, 42.10}
\vspace{2pc}
\noindent{\it Keywords}: Co-diffusion, complex contagions, synergy, multiplex networks modeling

\submitto{\NJP}
%\maketitle

\section{Introduction} \label{sec:introduction}

The term "social contagion" implicitly captures two worlds: the world of social science and the world of epidemiology. Although the term was initially coined in 1895 by Gustave Le Bon~\cite{le2017crowd} to describe undesirable collective behaviors in crowds, the definition has been stretched to encompass and explain types of collective behavior produced through social contact~\cite{bond201261}~\cite{wang2016dynamics}~\cite{wang2015dynamics}~\cite{ruan2015kinetics}~\cite{cozzo2013contact}. This broad definition, with advances in statistical physics~\cite{wang2016statistical}, has led Contagion Theory's inclusion within many avenues of social science research~\cite{guilbeault2017complex}, including marketing~\cite{van2004social}, innovation diffusion~\cite{iyengar2011opinion}, medicine~\cite{van2001medical}, health interventions~\cite{wang2017vaccination}, rumor and information spreading~\cite{moreno2004dynamics}~\cite{jalili2017information} sociology~\cite{burt1987social}~\cite{gallos2012people}, and the spread of emotion~\cite{fowler2008dynamic}~\cite{cacioppo2009alone}~\cite{hill2010emotions}.

In the same way that two contagions may influence each other's infectious paths, related innovations such as ideas, behaviors, products or technologies influence each others' diffusion. Prior social contagion models consider the spread of either one contagion at a time on interdependent networks or multiple contagions on single layer networks, usually under assumptions of pure competition. There is thus a desire to understand the diffusion of multiple social contagions under synergistic assumptions and to model the mechanisms for their concurrent, interfering spread. The paper has three main objectives. First, drawing upon established models within epidemiology and pharmacology, we propose a model which quantifies the amount of synergy between two contagions. Secondly, we consider the effects of stochastic dormancy, or “immunity,” towards similar contagions. This allows for the modeling of phenomenon that can be simultaneously cooperative and competitive. Lastly, we study the impact of network topology on diffusion, by contrasting short-range connections of lattice graphs and the long-range connections of a multiplex random-regular-graph.

The term "social contagion" is an oxymoron.  The word \textit{contagion} by itself means a disease spread by contact, a process in which the active and passive agents do not get to decide whether they are infected or not. On the other hand, social sciences are built on the assumption that rational individuals have the agency to make decisions. It is an exciting time for methodological intersection within computational social sciences, thus there is a desire to model agential decisions explicitly rather than stretching the interpretation of existing physical diffusion models.

The contemporary paradigm for diffusion research exists in network sciences, where a system is reduced into individual entities called nodes, each of which encapsulates characteristics of decision makers, which are then connected by edges. The most common critique of any mathematical representation is that they are too reductive. Thus, the greatest challenge is capturing "realistic" behaviors while keeping assumptions simple and tractable. This paper does so in two ways. First, it implements a multiplex network. The second, with a diffusion mechanism that encapsulates a simple definition of agency--- by relaxing the definition of immunity commonly found within the field of epidemiology, the spread of a contagion becomes a decision rather than a natural and neutral phenomenon.

We use the term social contagion broadly, pertaining to ideas, behaviors and a field of immediate application: innovation, technology, and product diffusion. The difference between competition and complementary action becomes difficult to identify when competition may, in fact, help another innovation. A new brand can 1) increase the entire market potential due to its promotion and variety or 2) compete for the same market potential and slow down the diffusion~\cite{chandrasekaran2007critical}. Given the role of social contagions in the process of innovation~\cite{langley2012determinants}, a quantification of synergy could frame this distinction in precise, numerical terms.

As an example, consider General Purpose Technologies (GPTs). GPTs denote technologies that have an impact across sectors and spillover effects as network externalities with economic benefits. Economists Lipsey and Carlaw have defined 24 GPTs, classified with four criteria~\cite{lipsey2005economic}:

\begin{enumerate} 
\item Is a single, recognizable generic technology
\item Initially has much scope for improvement but comes to be widely used across the economy
\item Has many different uses
\item Creates many spillover effects
\end{enumerate}

GPT's give rise to new products and innovations, many of which are complementary.  With the imminent arrival of the next GPT revolution with technologies such as artificial intelligence and blockchain, finding means to quantify synergy and thus predict the diffusion of related innovations is both timely and useful.

This paper is organized in the following way. Section~\ref{sec:lit-review} reviews the literature on diffusion. Section~\ref{sec:methods} develops necessary terminology and formulates the methodology. Section~\ref{sec:results} analyzes the results of the experiments. Section~\ref{sec:conclusion} concludes the study, examines the weaknesses, discusses interpretation within certain areas of the social sciences, then proposes future avenues of investigation. 

\subsection{Literature Review}  \label{sec:lit-review}

Guilbeault et al.~\cite{guilbeault2017complex} give three contemporary research directions. First is the ecology of contagions, which studies how \textit{multiple contagions} diffuse across different network structures. The second is the diffusion mechanism by modeling different threshold structures. The third is addresses structural diversity and examines the effects of homophily. This study focuses on the first two.

The first mathematical treatment of social contagion diffusion occurs with the Diffusion of Innovation based on the work of Rogers~\cite{rogers2010diffusion} and ~\cite{bass1969new}. Later, models within statistical physics~\cite{wang2016statistical} and epidemiology were appropriated for diffusion studies. Of particular relevance are models of multiple infections, where the increased likelihood of becoming infected conditional on a first due to weakened immune systems is well-documented. Nowak and May were one of the first to model super-infection, where they assumed only the strongest virus is active and thus the only one that spreads~\cite{nowak1994superinfection}. Shortly afterward, they modeled co-infection where multiple viruses are active~\cite{may1995coinfection}. Super-infection was then shown to be a limit of co-infection, and also gives the parameters in which multiple viruses can coexist. Similar to technology diffusion studies, competition is also modeled using cross-immunity in the context of a network~\cite{van2014domination}.

These models were then extended to study social contagions, particularly interacting contagions and antagonism~\cite{czaplicka2016competition}~\cite{velasquez2017interacting}~\cite{zhao2013percolation}, and their cascading behaviors on multiplex networks~\cite{brummitt2012multiplexity}~\cite{yaugan2012analysis}~\cite{lee2014threshold}~\cite{kivela2014multilayer}. Recently, Shu et al.~\cite{shu2017social} presents the dynamics of social contagions on two interdependent two-dimensional lattices. They give examples of nodes in communication networks which are spatially embedded among other applicable areas~\cite{barthelemy2011spatial}~\cite{daqing2011dimension}~\cite{boccaletti2014structure}. Li et al.~\cite{li2014epidemics} have a similar set-up to study the spread of epidemics, but in two experiments they first pair two interconnected lattices, then pair two Erdos-Renyi networks.  In contrast, this paper considers one lattice and one random-regular-graph, and thus investigates the interplay of spatial and long-range graphs. Aleta and Moreno~\cite{aleta2018multilayer} give a comprehensive review of how multilayer networks are used various contexts, including diffusion and percolation, and how this is applied to ecology, biology, transportation, economics, game theory, and transportation. 

Another approach towards diffusion on networks is with evolutionary games~\cite{zinoviev2011game}~\cite{qiu2012game}. Jiang ~\cite{jiang2014graphical} is one of the first to specifically treat the mechanism to actively spread information using game theory, and apply it to social networks. New information is treated as a mutant, and there are recent efforts to extending this to multilayer networks~\cite{jiang2014graphical}~\cite{perc2013evolutionary}. Factors of importance within networks include noise-induced adoption and network topology~\cite{perc2006double}, collective influence by degree~\cite{szolnoki2016collective}, teaching activity~\cite{szolnoki2008coevolution}, and stochasticity~\cite{perc2006evolutionary}. Perc's treatment of noise is of particular relevance given the well-documented spontaneous adoption of behavior~\cite{hill2010infectious}

For the diffusion mechanism, one common paradigm is the susceptible-infected-susceptible (SIS) model within network~\cite{ball1997epidemics}~\cite{watts1998collective}~\cite{keeling1999effects}. In the basic SIS model, individuals transition between states of susceptibility and infection, where recovered individuals are once more susceptible. In other words, recovery from disease confers no immunity. Hill et al.~\cite{hill2010infectious} characterize a SISa model, in which they distinguish between spontaneous and contagious infection. Hill et al. also mentions many of the challenges this paper wishes to address: probabilistic nature of the contagion and asymmetry within the contagions. Hill et al. ~\cite{hill2010emotions} have also modeled emotion as an infectious disease in large-scale networks using the same SISa model. Dodds and Watts~\cite{dodds2005generalized} provide a generalized model for social and biological contagions also using the SIS model, and identify three basic classes of contagion models which they call \textit{epidemic threshold}, \textit{vanishing critical mass}, and \textit{critical mass}, and how one may interpret results for prevention or facilitation based on these cases. While the examples above occur on single layer networks, Liu et al.~\cite{liu2015epidemics} have modeled epidemic spread on interconnected small-world networks, where neighbors of a node are likely to be neighbors of other nodes. Precisely, the typical distance $L$ between two nodes chosen at random will grow proportionally to $\log N$ with $N$ the number of nodes. In regards to multiple infections, Chen et al.~\cite{chen2017fundamental} propose a model built on intrinsic properties of “cooperative” contagions $A$ and $B$. Their model is also based on the SIS model where host individuals are in two possible states: susceptible (S) or infected(I). Susceptibles are equivalent to naive agents and can be infected by either $A$ or $B$, each associated with a cooperativity coefficient $\xi_A$ and $\xi_B$ respectively to capture their mutual influence. 

The Susceptible-Infection-Recovered (SIR) model, in contrast, confers a removed or recovered status to individuals, who are no longer susceptible to disease~\cite{nowak2006evolutionary}~\cite{anderson1992infectious}. Immunity is a parameter that has analogous application in social systems, such as resistance to rumors-spreading or belief-change. However, immunity is bidirectional. Gaining recovered status precludes the node from being \textit{both} infected and from infecting others, a distinction of great import to our study.

\section{Materials and Methods} \label{sec:methods}

\subsection{Definitions and Terminology}
Networks are represented mathematically by graphs. A graph is notated as $G=(V,E)$, where $V$ is a set of nodes called vertices and $E$ is the set of links called edges. Every element in $E$ is represented by the Cartesian Product of two vertices $V \times V = (v_i, v_j)$, since each link must have two endpoints. If $(v_i,v_j)$ is ordered, the graph is said to be directed; if order is irrelevant such that $(v_i,v_j) = (v_j,v_i)$ then the graph is undirected. A multiplex network is a network where the same set of nodes are represented in every layer, although the interactions and links between the nodes may be different. 
We use multiplex networks to differentiate between short-range and long-range connections for different dynamical channels or modes~\cite{kivela2014multilayer}.

\subsection{The General Algorithm}
% It is tempting when working with mathematical models to introduce parameters that may be only meaningful within the circle of applied mathematics. Thus, the variables and framework included are justified with fundamental and well-established facts from social contagion theory and innovation diffusion theory, and utilize analogous aspects of existing epidemiological models.

Without loss of generality, assume that we have two contagions: Contagion \emph{A} and Contagion \emph{B}. Every node in the experiment must take on one of four states: naIIIve ($\emptyset$), $A$, $B$, and $AB$, which correspond to uninfected/naive individuals, adopters of Contagion $A$, adopters of Contagion $B$, and adopters of both Contagion $A$ and $B$. Diffusion denotes the spread of a contagion across a network, which influences the particular status of a node. Furthermore, each node is either active or dormant, represented by the Boolean $1$ or $0$ respectively. The status for each node is expressed as the tuple $(\textit{State}, \textit{Activity})$. For instance, an active node infected with $A$ would be represented by $(A,1)$. Updating is synchronous~\cite{schonfisch1999synchronous}. Each iteration of the simulation can be summarized as:
\begin{itemize}
\item Initialize a graph
\item Seed Contagion A and Contagion B
\item Diffuse graph by one time-step
\item Count the nodes for the Uninfected, Contagion A, Contagion B, and both A and B
\item Repeat from Step 3 until the last time-step
\end{itemize}

The raw output for each experiment are four time series. The general algorithm is described in Algorithm~\ref{alg:diffusion}:

\begin{algorithm} \label{alg:diffusion}
\caption{Algorithm for simulating network diffusion}
\label{alg:diffusing-graph}
$\alpha$-list = list of alpha values \\
$\tau_A$-list = list of $\tau_A$ values \\
$\tau_B$-list = list of $\tau_B$ values, with the constraint that $\TB<\TA$ \\

$\textit{ParameterList}$ = list of tuples containing $(\alpha, \tau_A, \tau_B)$ \\

\For{ParameterSet in ParameterList}{
\For{Number of Iterations}{
Create graph $G$\\
Randomly seed one of Contagion A and Contagion B into the nodes, non-overlapping\\
Set a random threshold for each node\\
\For{All time-steps}{
Diffuse the Contagions using a multivariate probability distribution \\
Deactivate nodes based on probabilities $\tau_A$ and $\tau_B$\\
Count the total number of Uninfected, Contagion A, Contagion B, and Both Contagions
}
Save the individual time-series
}
Average time-series for each of the four categories
}
\end{algorithm}

The full list of parameters can be found in Section~\ref{sec:variables}. Simulations were parallelized across parameters and run using a cluster of 600 CPUs. The sections below walk through the specific details of each step in Algorithm~\ref{alg:diffusing-graph}.

\subsection{Initializing the Graph}
\begin{figure}[!htb]
        \centering
        \includegraphics[width=1.0\linewidth]{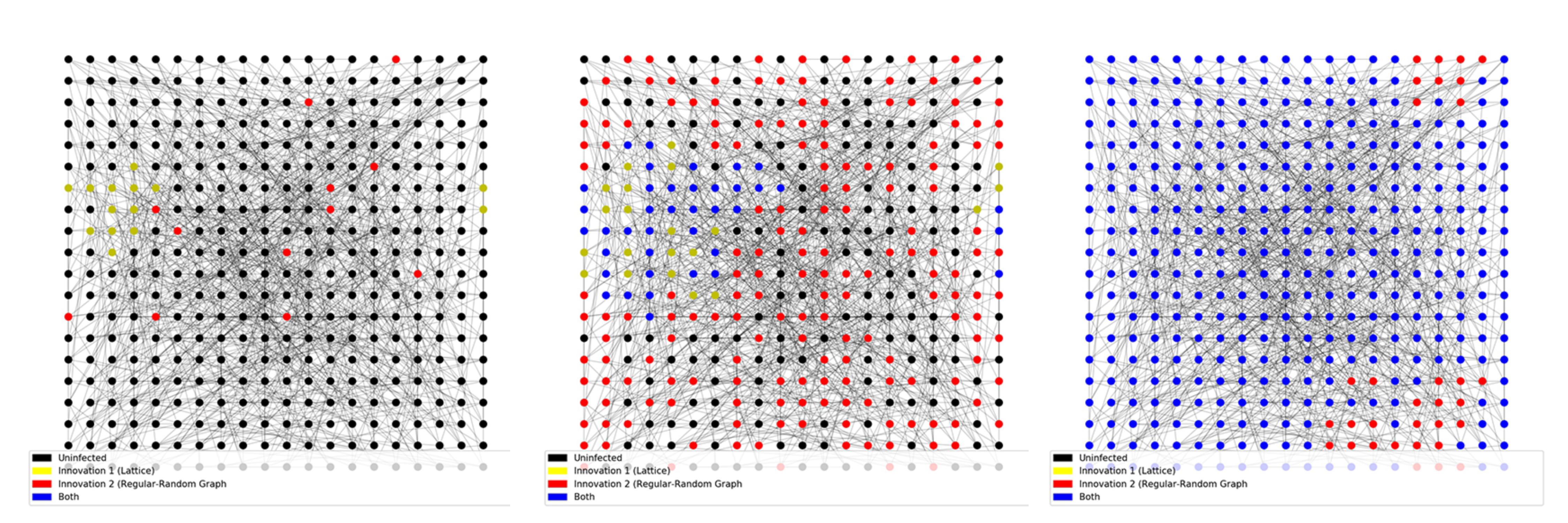}
        \caption{Multiplex Diffusion}{Contagion A (yellow) spreads across the periodic lattice. Contagion B (red) spreads on the random-regular-graph, which show up as isolated points on the lattice. The dark, faded blotches are the edges of the random-regular-graph.}
        \label{fig:multiplex-diffusion}
\end{figure}

The diffusion simulation occurs on a multilayer network. One layer is the periodic square lattice, the second is a random-regular-graphs with degree four. The distinction between the periodic square lattice and the random-regular-graph is that now long-range edges are allowed, so we can compare the diffusion results of long-range connections versus the spatially localized.  As we will discuss in Section~\ref{sec:methods-threshold-kernel}, the threshold for $A$ is determined by the neighbors on the lattice graph while the threshold for $B$ is determined from the random-regular-graph. Figure~\ref{fig:multiplex-diffusion} illustrates the diffusion process.

From these nodes, one node is randomly selected, then seeded with Contagion A. Another one is seeded with Contagion B, with no repetition.

% \begin{figure}[!htb]
%         \centering
%         \includegraphics[width=1.0\linewidth]{Methods/lattice-diffusion.jpg}
%         \caption{Single-Layer Lattice Diffusion}{Innovation A (yellow) and Innovation B (red) spread to their von Neumann neighbors, then overlap with increased time steps as shown in blue.}
%         \label{fig:lattice-diffusion}
% \end{figure}

For each node at each time step, its activity and status (denoted as Uninfected, $A$, $B$, or $AB$) is updated. The following table summarizes the experimental parameters:
\begin{table}[!htb]
\centering
\caption{Exogenous Variables}
\label{tab:exo-var}
\begin{tabular}{@{}ll@{}}
\toprule
\textbf{Parameter}           & \textbf{Quantity} \\ \midrule
Number of Nodes              & $6400$            \\
Total Timesteps              & $700$             \\
Number of Initial Seeds      & $1$               \\
Iterations per Parameter Set & $100$             \\ \bottomrule
\end{tabular}
\end{table}

\subsection{Adoption Probability Kernel}  \label{sec:methods-adoption-probability}
Section~\ref{sec:lit-review} of the literature review describes the assumptions within innovation diffusion, one of the main ones being that we can model the diffusion using a logistic function. The Hill Function is the log-transform of the logistic function and is useful in modeling density-dependent growth since it takes density as a direct parameter. It has roots in biochemistry is was used to measure the rate of reaction between reactant concentration and substrate density. The Hill function is given as:

\begin{equation} \label{eq:hill}
f(X) =  \frac{[X]^\alpha}{[X]^\alpha + K^\alpha}
\end{equation}

where $[X]$ is the density of X, $K$ denotes the time-step that is half-way to full saturation, and $\alpha$ is a shape parameter that determines how steep the slope of the function is. In other words, increasing $K$ corresponds to shift right, although its shape becomes less steep as well. The ratio between $\alpha$ and $K$ influences the steepness.  Researchers in pharmacology noticed that the effect of drugs on killing cells and bacteria could be modeled similarly, and then extended the model to capture the effects of drug interaction.
\subsubsection{The BLISS and Loewe Additivity} \label{sec:loewe}

Traditional Chinese medicines typically use mixtures of herbs to maximize the potency during the healing process. This notion has not escaped the contemporary field of pharmacology, and the last century has seen the increased use of drug combinations.

Since studies have shown that for social contagions, density matters more than numbers, existing models within the pharmacology literature prove suitable for the co-diffusion model. Theoretical research within the field quantify the concepts of synergy and antagonism, which go beyond the simple additive effect of using drugs individually.

The Bliss Independence Model~\cite{foucquier2015analysis} assumes that drug effects are outcomes of probabilistic processes, but each contributes to a common result. This model is filed under effect-based strategies, which compare the effects resulting from both drugs versus individual components. In this case, effect refers to the efficacy of killing bacteria or certain types of cells.

The second strategy for analyzing the efficacy of two drugs considers what concentration of each produces the same response~\cite{foucquier2015analysis}. The mathematical framework known as Loewe Additivity utilizes this Dose Equivalence Principle (DEP) to formalize definitions of synergy, additivism, and antagonism. The DEP states that dose $a$ is equivalent to $b_a$ where~\cite{foucquier2015analysis}
$$
E(a+b) = E(a+a_b) = E(b_a+b)
$$
For the sake of discussion in this section, suppose we have drugs A and B, each administered at doses $a$ and $b$ respectively. This yields the combination index
$$
Cwe = \frac{a}{A} + \frac{b}{B}
$$.
For $CI<1$, this indicates that that the combined amount of the two drugs required to produce a certain effect is less than the amount required when they are used individually. In other words, the combination is more effective and thus they act synergistically. Similarly, $CI>1$ indicates that the combination of $a$ and $b$ produces a worse effect and is antagonistic.

Dose effects typically follow the logistic shape of the Hill equation defined as:
$$
E = E_{max} \frac{a^\alpha}{k_a + a^\alpha}
$$
where $a$ denotes the concentration of $A$, $E_{max}$ denotes the maximum effect or the ceiling, and $k$ is the inflection point of the curve. What this model gives us is a general framework that is grounded in similar empirical observations in the fields of innovation diffusion and drug effectiveness. Instead of considering how two drugs work together to kill cells, we consider how innovations spread concurrently.

\subsubsection{Probability Kernel for the Threshold} \label{sec:methods-threshold-kernel}

We treat an increase in the number of neighbors as a \textbf{threshold lowering effect}. As with the typical threshold model, each node is assigned a threshold $\mu_k$, where:
$$
\mu_i \in (0,1) \qquad \textit{where } i \textit{ is the coordinate of each unique node }
$$
A simple threshold can be constructed with Equation~\ref{eq:hill}, but it is insufficient to characterize the logistic shape of diffusion in different network settings. Here we develop thresholds under the assumption that each innovation individually diffuses according to the shape of a Hill Function.
We have assumed, without loss of generality, that we have two products \emph{A} and \emph{B}. Assume that inclusive adoption is possible. We also assume that only one innovation can be adopted per time-step. Then there are two possible paths of adopting both \emph{A} and \emph{B}. Let $i$ denote a node.
\begin{equation}
\begin{aligned}
i(\textit{Naive}) \rightarrow i(A) \rightarrow i(AB) \\
i (\textit{Naive}) \rightarrow i (B) \rightarrow i (AB)
\end{aligned}
\end{equation}
or simply put, the naive/uninfected individual must adopt A first or B first. This is known as inclusive adoption. Exclusive adoption denotes the case where the state $\phi_i (AB)$ is not possible.
Next, we denote an indicator function for the status where:

$$
\begin{aligned}
S_A(i) = 
\begin{cases}
1   & \textit{if } i \textit{ adopts A} \\
0  & \textit{if otherwise}
\end{cases}  \qquad \qquad
S_B(i) = 
\begin{cases}
1   & \textit{if } i \textit{ adopts B} \\
0  & \textit{if otherwise}
\end{cases}
\end{aligned}
$$

Thus, the inclusive adoption probability of any state can be expressed using this general formula:
\begin{equation} \label{eq:adoption-general}
P(i) = \frac{ \Big( 1- S_A(i)  \Big) \Big( \frac{[A]}{K_A} \Big)^\alpha   + 
\Big( 1- S_B(i)  \Big) \Big( \frac{[B]}{K_B} \Big)^\alpha  }{1 + \Big( 1- S_A(i)  \Big) \Big( \frac{[A]}{K_A} \Big)^\alpha   +  \Big( 1- S_B(i)  \Big) \Big( \frac{[B]}{K_B} \Big)^\alpha  }
\end{equation}
where $K_A$ and $K_B$ controls the attractiveness of each social contagion. The smaller the value, the more attractive it is to the population since it controls for the time step of the inflection point. $[A]$ denotes the density of neighbor nodes that have already adopted innovation $A$. Specifically, let $T$ be the total number of neighbors, then:
\begin{equation}
[A] = \frac{\textit{No. of A}}{T}
\end{equation}
The assumption is that $\alpha$ and $K$ are known and can be fit based on past data. For the purpose of this study we assume $K_A = K_B = 2.0$, where the choice of this parameter suits the simulation time scale.
To clarify Equation~\ref{eq:adoption-general}, we break down the sub-cases. For the naive individual $i$ the values of $S_A(i)$ and $S_B(i)$ are both zero. Hence the adoption rate of either $A$ or $B$ can be characterized by:
\begin{equation} \label{eq:naive-adoption}
P(i \leftarrow A \textit{ or } B  ) = \frac{ \Big( \frac{[A]}{K_A} \Big)^\alpha   +\Big( \frac{[B]}{K_B} \Big)^\alpha  }{1 + \Big( \frac{[A]}{K_A} \Big)^\alpha   +\Big( \frac{[B]}{K_B} \Big)^\alpha  }
\end{equation}

The node is first activated with this probability. Then it will choose one of $A$ and $B$ based on their relative proportions. That is, 
\begin{equation} \label{eq:A-B-coinflip}
\Pr (i \leftarrow A) = \frac{ \Big( \frac{[A]}{K_A} \Big)^\alpha  } { \Big( \frac{[A]}{K_A} \Big)^\alpha   +\Big( \frac{[B]}{K_B} \Big)^\alpha  } \qquad
\Pr (i \leftarrow B) = \frac{ \Big( \frac{[B]}{K_B} \Big)^\alpha  } { \Big( \frac{[A]}{K_A} \Big)^\alpha   +\Big( \frac{[B]}{K_B} \Big)^\alpha  }
\end{equation}

\begin{figure}[!htb]
        \centering
        \includegraphics[width=1.0\linewidth]{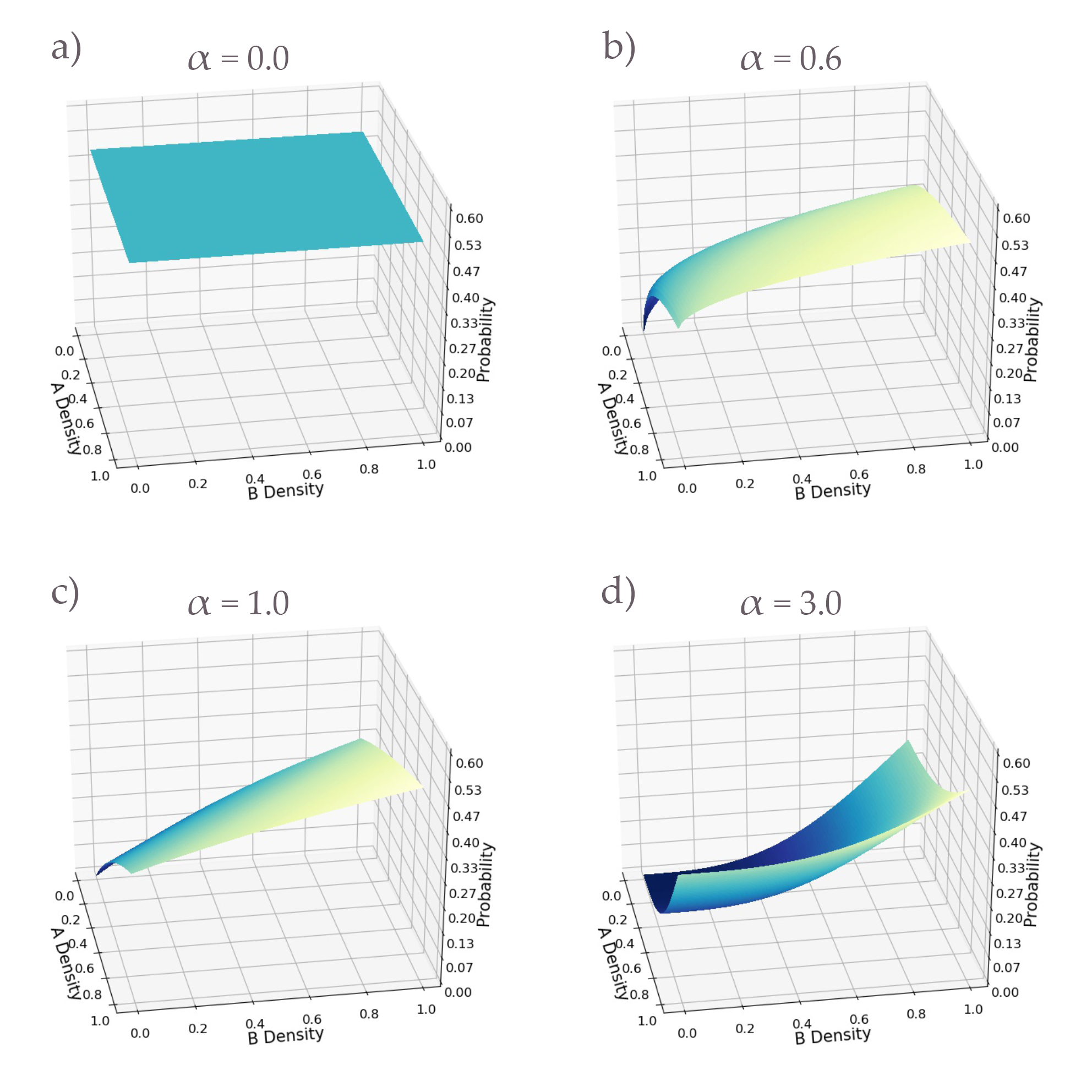}
        \caption{Adoption Probability Kernel}{The surface of the multivariate Hill Function (Equation~\ref{eq:naive-adoption}), which describes the probability of adopting either A or B, depending on the densities of $A$ and $B$. Here we assume $K_A = K_B = 2.0$ such that the parametrization is symmetric between Contagion $A$ and Contagion $B$, though the parameter can be adjusted for asymmetry. Diagram a) denotes constant adoption, b) denotes synergestic additivity, c) denotes near linear additivity, and d) denotes antagonistic additivity.  This adoption kernel is used to study diffusion on a multiplex graph of 6400 nodes, with a lattice and a Random Regular Graph of degree 4.}
        \label{fig:adoption-surf}
\end{figure}

Here, the notation $\Pr (i \leftarrow A)$ denotes the probability of node $i$ adopting Contagion $A$, and it is the analogous case for $B$. The adoption probability is shown as a surface in Figure~\ref{fig:adoption-surf}. In lieu of Loewe Additivity described in Section~\ref{sec:loewe}, these surfaces indicate the different relationships of synergy and antagonism. When $0 <\alpha < 1.0$, the curve is concave downwards, which indicates that the effect of their sum is more than their parts and thus synergistic. When $\alpha > 1.0$ as with the case in the bottom right, the relationship is concave upwards which indicates antagonism. This corresponds to the formulation given in the isobologram analysis of ~\cite{foucquier2015analysis}. 

Now without loss of generality suppose $i$ has already adopted $A$. Then the probability of adopting $B$ is given by
$$
\frac{\Big( \frac{[B]}{K_B} \Big)^\alpha }{1+\Big( \frac{[B]}{K_B} \Big)^\alpha }
$$
The case for adopting $A$ after first adopting $B$ is analogous.

In the experiment we set the thresholds as a function of this adoption probability. The threshold $\mu_i$ is given in Equation~\ref{eq:threshold}:
\begin{equation} \label{eq:threshold}
\mu_i  = 1-P(i \leftarrow A)
\end{equation}

Then at every time step, a random number is chosen between 0 and 1 to determine the probability of adoption. Ultimately, the formulation shown in Equation~\ref{eq:adoption-general} captures synergistic diffusion as the initial adoption depends on both densities.

\subsubsection{Exclusive Adoption} \label{sec:excl-adoption}
While not explicitly studied in this paper, exclusive adoption is a useful contrast to our case above. The adoption pathways can be represented as:
\begin{equation}
\begin{aligned}
i(\textit{Naive}) \rightarrow i(A) \rightarrow \textit{Immunity against B} \\
i (\textit{Naive}) \rightarrow i (B) \rightarrow \textit{Immunity against A}
\end{aligned}
\end{equation}
The expression for adopting any of the contagions is thus:
\begin{equation} \label{eq:naive-adoption-exclusive}
P(i \leftarrow A \textit{ or } B  ) = \Big( 1 - S_A(i) \Big) \Big( 1 - S_B(i) \Big)
\frac{ \Big( \frac{[A]}{K_A} \Big)^\alpha   +\Big( \frac{[B]}{K_B} \Big)^\alpha  }{1 + \Big( \frac{[A]}{K_A} \Big)^\alpha   +\Big( \frac{[B]}{K_B} \Big)^\alpha  }
\end{equation}
It then chooses $A$ or $B$ with the coin-flip expressed in Equation~\ref{eq:A-B-coinflip}.

\subsection{Stochastic Dormancy} \label{sec:stochastic-dormancy}
The prior two parameters $\alpha$ and $K$ model the shape of diffusion, and thus they only influences the timescale of diffusion. As time approaches infinity, the diffusion process will always diffuse to the maximal value. This is not the case in reality, as the penetration depth is usually a subset. We model this by introducing stochastic dormancy to every node on the graph, such that nodes are not active in perpetuity.

To do this, we attach a constant $\tau_A$ and $\tau_B$ to contagions A and B respectively. $\tau_A$ denotes the probability that a node infected with $A$ will become dormant at any given time step. When a node is considering adoption, if a neighbor is dormant then that neighbor is discounted from the numerator of the density, and is thus not included in the threshold lowering effect. To be numerically precise,
\begin{equation}
[A] = \frac{\textit{No. of Active A}}{T}
\end{equation}
The same holds for Contagion B. Another way of interpreting $\tau$ is that $\tau_A$ represents the average proportion of nodes infected by $A$ that will switch off at each time-step. For nodes infected with both $A$ and $B$, the $\tau_{AB}$ value is simply the arithmetic average; different conditions for convexity is another line of inquiry. The expression is given as:
$$
\tau_{AB}  = \frac{\tau_A + \tau_B}{2}
$$
% As discussed in Section~\ref{sec:epidemiology}, 

It is important to distinguish between \textit{immunity} and \textit{dormancy}. 
In epidemiology both immunity and recovery imply two things--- a recovered individual can no longer be infected nor can it infect other nodes. For the purpose of studying social contagions, we relax the first condition. In other words, while inactive individuals no longer affect other nodes, they themselves can still be infected by another contagion. Thus, they do not gain immunity from being infected, but they do less to help the contagion spread. We make a basic assumption that neighbors have to make a decision and play an active role in the diffusion process.
 
\subsection{Experimental and Measured Variables} \label{sec:variables}

\begin{table}[!htb]
\centering
\caption{Independent Variables}
\label{tab:ind-var}
\begin{tabular}{@{}ll@{}}
\toprule
\textbf{Parameter}         & \textbf{Quantity} \\ \midrule
$\alpha$ range             & $0.0$ to $1.3$    \\
$\tau$ Range for Multiplex & $0.0$ to $1.0$    \\ \bottomrule
\end{tabular}
\end{table}

\begin{figure}[!htb]
        \centering
        \includegraphics[width=0.6\linewidth]{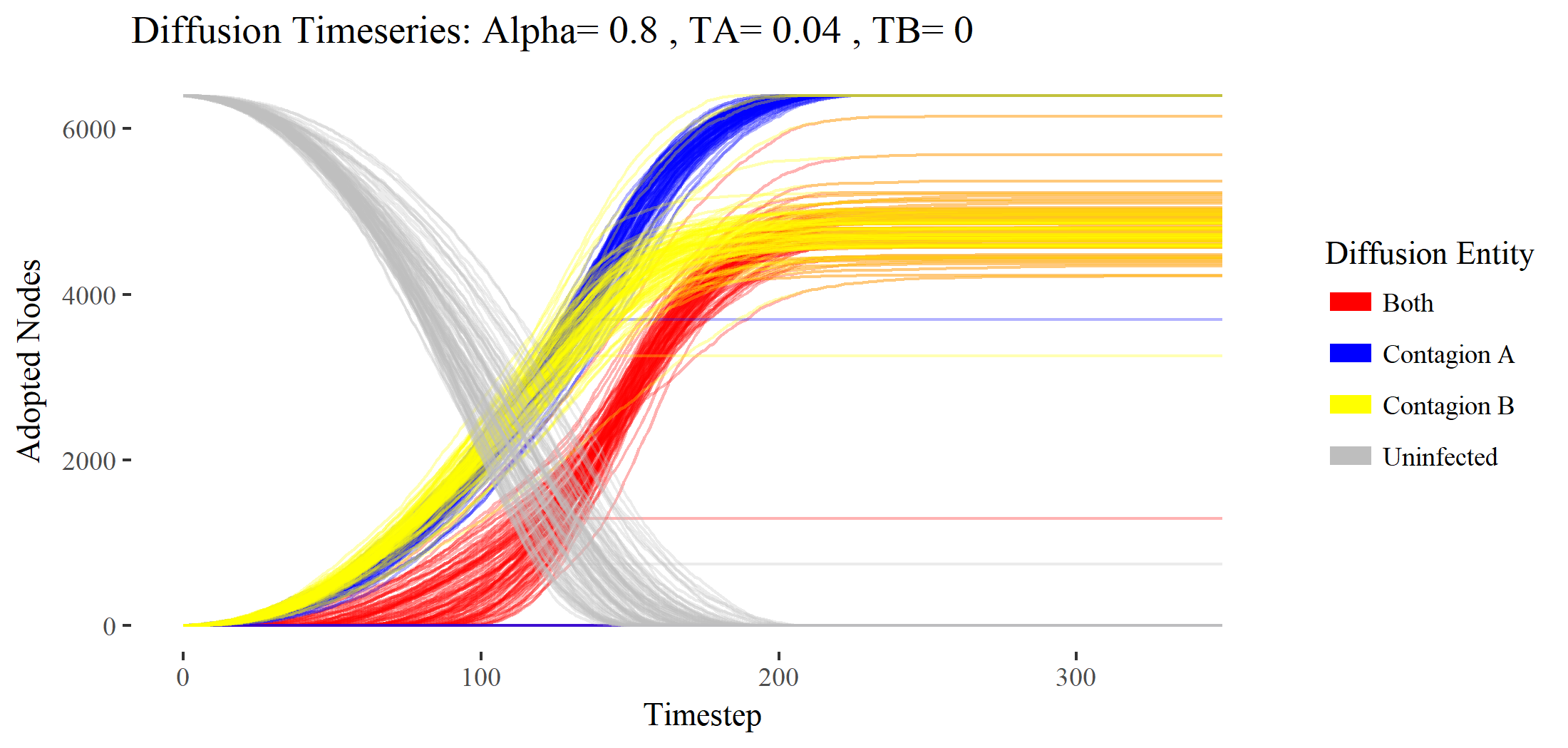}
        \caption{Time-Series of One Parameter Set}{ The diffusion curves of Contagion $A$ (blue), Contagion $B$ (yellow), Both $A$ and $B$ (red), and Uninfected nodes (grey) on a single layer lattice. Parameters are set with $\alpha = 0.8$, $\TA=0.04$, and $\TB=0.00$. }
        \label{fig:one-timeseries}
\end{figure}

The primary variable of interest is the depth of diffusion. This is taken as the equilibrium value at the end of diffusion, found by taking the diffusion depth of the final 20\% of periods. We use the term \textit{ceiling} in lieu of penetration depth. Secondly, we consider the inflection point as a proxy for rate, as it represents the time-step where the diffusion curve attains exactly half of its ceiling. If this point is pushed right, this indicates diffusion takes a longer time and thus the rate is lower. Figure~\ref{fig:one-timeseries} shows the output of 100 iterations of one set of parameters.

The diffusion on multiplex networks follows the same processes as Sections~\ref{sec:methods-threshold-kernel} and ~\ref{sec:stochastic-dormancy}, however, the network topology is a factor in counting the neighbors. Contagion $A$ diffuses on a $80 \times 80$ periodic square lattice and Contagion $B$ diffuses on a random-regular-graph with degree 4. For each node considering the adoption of Contagion $A$, only the active neighbors on the lattice graph are considered; for the adoption of Contagion $B$, only active neighbors on the random-regular-graph is considered. 

\section{Results} \label{sec:results}

\begin{figure}[!htb]
        \centering
        \includegraphics[width=1.0\linewidth]{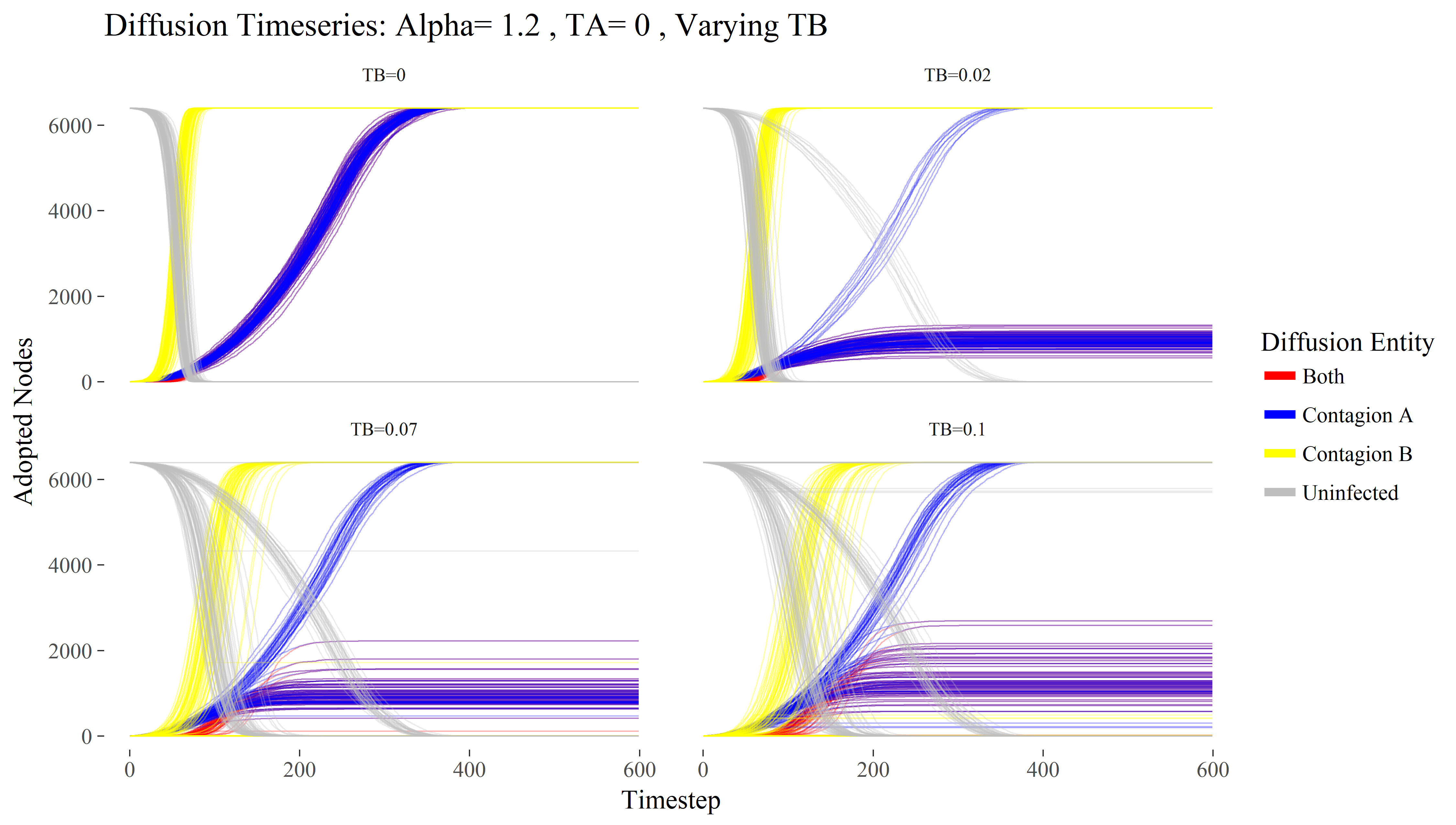}
        \caption{Multiplex Diffusion Curves while Varying $\tau_B$, fixing $\alpha=1.2$ and $\TA = 0$}{Introducing dormancy through $\TB$ produces branching in the diffusion curves. Small values of $\TB$ cause a "thicker" lower branch as shown in b). As $\TB$ increases to $0.1$, there is a higher likelihood Contagion $B$ itself undergoes percolation and thus the branches are more even.}
        \label{fig:multiplex-diffusion-TB}
\end{figure}

We begin the analysis by considering the diffusion curves directly. In Figure~\ref{fig:multiplex-diffusion-TB}, we immediately observe Contagion $B$ shown in yellow diffuses much faster than Contagion $A$. Given that $K$ are equal, this can be attributed to two advantages of long-range connections. First, as the diffusion process starts the uninfected nodes that Contagion $B$ is in contact with is much greater than that of Contagion $A$. The uninfected nodes that Contagion $A$ affects are always restricted to some physical front as seen in the left-most figure of ~\ref{fig:multiplex-diffusion}, like the propagation of a wave. Secondly, the long-range connections reduce the effect of percolation produced by $\TB$. In contrast, restricting diffusion to von Neumann neighborhood encourages local percolation since many of the naive individuals share neighbors. This increases the chance that a mutual neighbor no longer participates in diffusion and thus decreasing the density-based adoption probability on the overall front.
 
The central topic of this study is how two contagions can spread synergistically, and how their intrinsic properties affect one another. What is of great interest is the branching of Contagion A shown in blue in Figure~\ref{fig:multiplex-diffusion-TB}, where we fix the values of $\alpha=1.2$ and $\TA=0$ while varying $TB$. When Contagion B’s dormancy rate is increased marginally from $\TB = 0$ to $\TB =0.02$, a large degree of branching in Contagion A is observed, shown in Figure~\ref{fig:multiplex-diffusion-TB}b. In other words, the intrinsic dormancy of Contagion $B$ greatly influences the diffusion of $A$. As $\TB$ is increased from $0.02$ to $0.07$, a greater degree of blue curves converge maximally, and even more so when $\TB=0.10$.

\begin{figure}[!htb]
        \centering
        \includegraphics[width=1.0\linewidth]{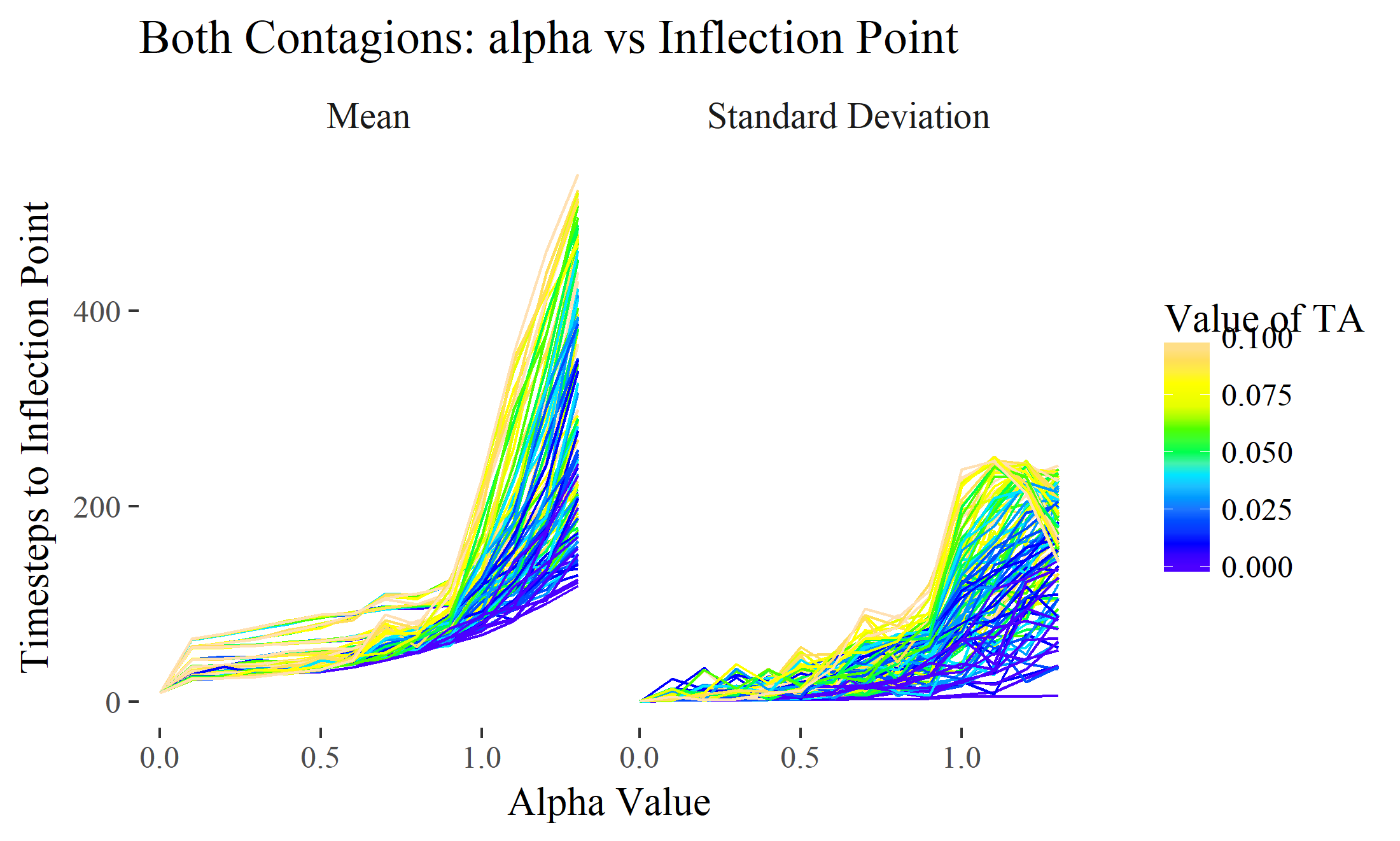}
        \caption{$\alpha$ slows diffusion for Multiplex graphs. }{}
        \label{fig:multiplex-rateVSalpha}
\end{figure}
 
To understand this instability, first consider the effect of synergy on the rate of diffusion. Figure~\ref{fig:multiplex-rateVSalpha} shows an increase in $\alpha$ slows down the rate of diffusion. Furthermore, there is a sharp rise in the rate of diffusion at around $\alpha=0.8$, where the adoption kernel shown by Equation~\ref{eq:adoption-general} and Figure~\ref{fig:adoption-surf} transitions from synergistically additive to antagonistically additive. Additionally, note that as $\TA$ increases, the rate of diffusion also increases. Since this is the diffusion of nodes infected with both $A$ and $B$, and since $A$ diffuses more slowly, this implies the dormancy of the contagion diffusing more slowly produces percolation.
 
\begin{figure}[!htb]
        \centering
        \includegraphics[width=1.0\linewidth]{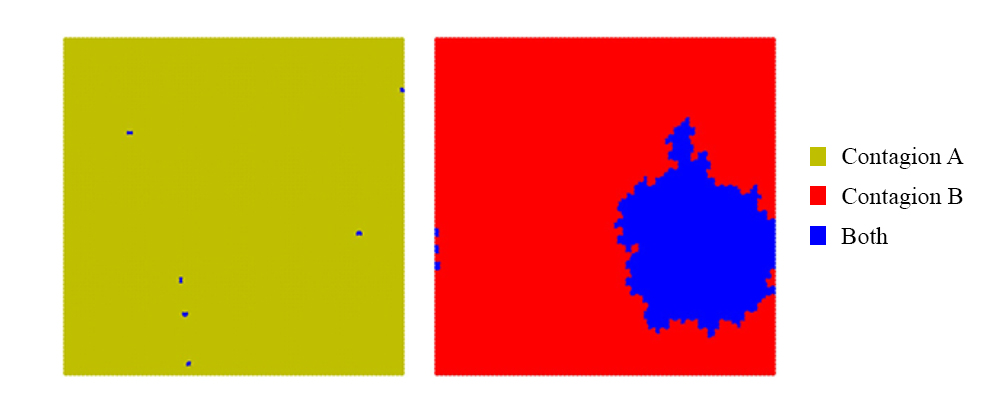}
        \caption{Convergent outcomes from the branching. Contagion $A$ diffuses completely on the left as Contagion $B$ has stopped diffusion due to $\TB=0.1$. On the right, Contagion $B$ diffuses fully and produces ring vaccination against $A$. Not shown is a sub-case of the left, where Contagion $B$ can diffuses fully with a "synergestic boost" from Contagion $A$.}{}
        \label{fig:branched-outcomes}
\end{figure}
 
Figure~\ref{fig:branched-outcomes} shows the two possible outcomes produced from the branching, though not the only two. Figure~\ref{fig:branched-outcomes}a occurs when Contagion $B$ is stopped by its own dormancy rate, thus allowing Contagion $A$ to fully diffuse to its von Neumann neighbors and across the entire lattice. Figure~\ref{fig:branched-outcomes}b, on the other hand, occurs when Contagion $B$ cascades and diffuses first, and over time introduces dormancy into the population. Notice that Figure~\ref{fig:branched-outcomes} b converges to equilibrium much faster, which is reflected in the inflection points of Figure~\ref{fig:multiplex-diffusion-TB}. 
A deep result is the ring vaccination shown in Figure~\ref{fig:branched-outcomes}b, where dormancy in the surrounding neighbors percolates the diffusion of Contagion A, only requires this one directional definition of dormancy. In other words, resistance is potentially unnecessary for ring vaccination given sufficient lack of spreading.

We are now ready to summarize the interaction of $\tau_A$ and $\TB$  using heatmaps; we begin with Contagion $A$.

\begin{figure}[!htb]
        \centering
        \includegraphics[width=1.0\linewidth]{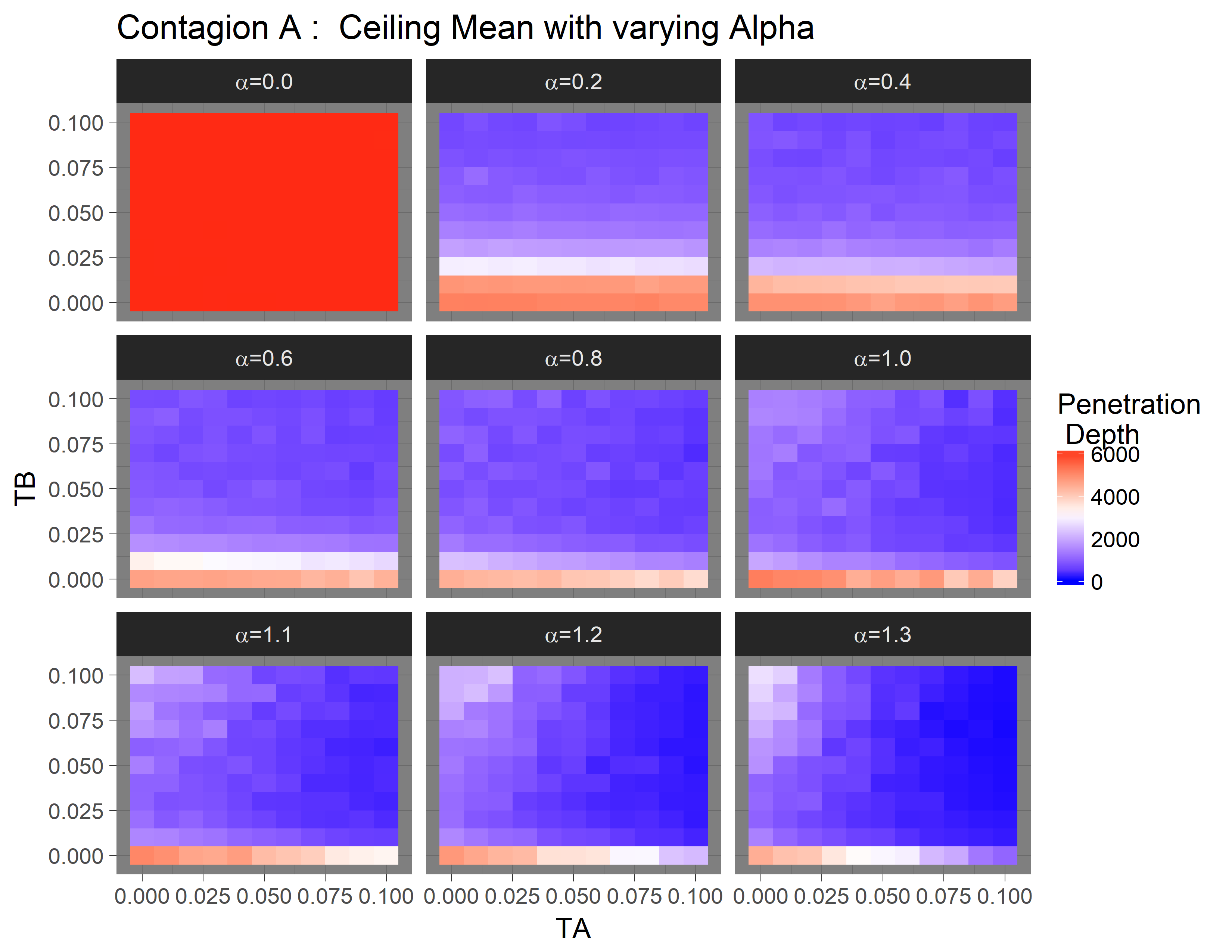}
        \caption{Ceiling Mean of Contagion A for Multiplex Experiment}{The ceiling of $A$ demonstrates high sensitivity to $\TB$ when $\alpha<0.8$. It shows sensitivity to $\TA$ when $\alpha>1.0$.}
        \label{fig:multiplex-A-C-mean}
\end{figure}
 
When $\alpha=0$ the diffusion probability is uniformly $0.5$ as shown in the top left of Figure~\ref{fig:adoption-surf}. Diffusion is both constant and rapid which forces Contagion $A$ to converge maximally. This is similar to the case of single layer diffusion. As $\alpha$ increases and diffusion slows, $\TB$ has a very pronounced effect on the ceiling. Consider the case when $\alpha = 0.2$ and $\TB$ jumps from $0.02$ to $0.03$. The mean ceiling diminishes rapidly shown by the color change of red to white. As $\alpha$ increases the decrease in ceiling mean is even more pronounced, which produces the very noticeable color jump between $\TB=0$ and $\TB=0.01$. This rapid drop in the ceiling is a testament to how much faster $\TB$ diffuses, where even low values of $\TB$ have sufficient time to inject dormancy into the populace.
Consider the last row of the grid where $\alpha=1.1,~1.2$ and $1.3$. When we hold each $\TB$ constant we observe a gradient effect left to right from increasing $\TA$. This is most pronounced when $\TB=0$. We conclude that if $\alpha>1.0$, then $\TA$ has a large effect on the ceiling of Contagion $A$.

The previous two observations on $\TB$ and $\TA$ respectively allow us to conclude that increasing $\alpha$ compresses the graph from the right and from the top. The two red horizontal rows in the $\alpha = 0.2$ grid are eventually compressed into one row from above. The horizontal gradient of $\TB=0$ and $\alpha=1.1,~1.2$ and $1.3$ (bottom row of the grid) is compressed towards the left. Increasing $\alpha$ decreases the synergy, slows diffusion, and thus gives $\tau_A$ and $\tau_B$ more time to produce dormancy although they demonstrate different sensitivities towards $\alpha$.

When $\TB$ is high the ceiling of $A$ actually increases. This phenomenon is due to $\TB$ slowing the diffusion of Contagion $B$, which produces more branched "upper" curves of Contagion $A$, and we can see this in Figure~\ref{fig:multiplex-diffusion-TB} by comparing $\TB = 0.02$ and $\TB=0.1$. This can also be shown with the standard deviation heatmap.

\begin{figure}[!htb]
        \centering
        \includegraphics[width=1.0\linewidth]{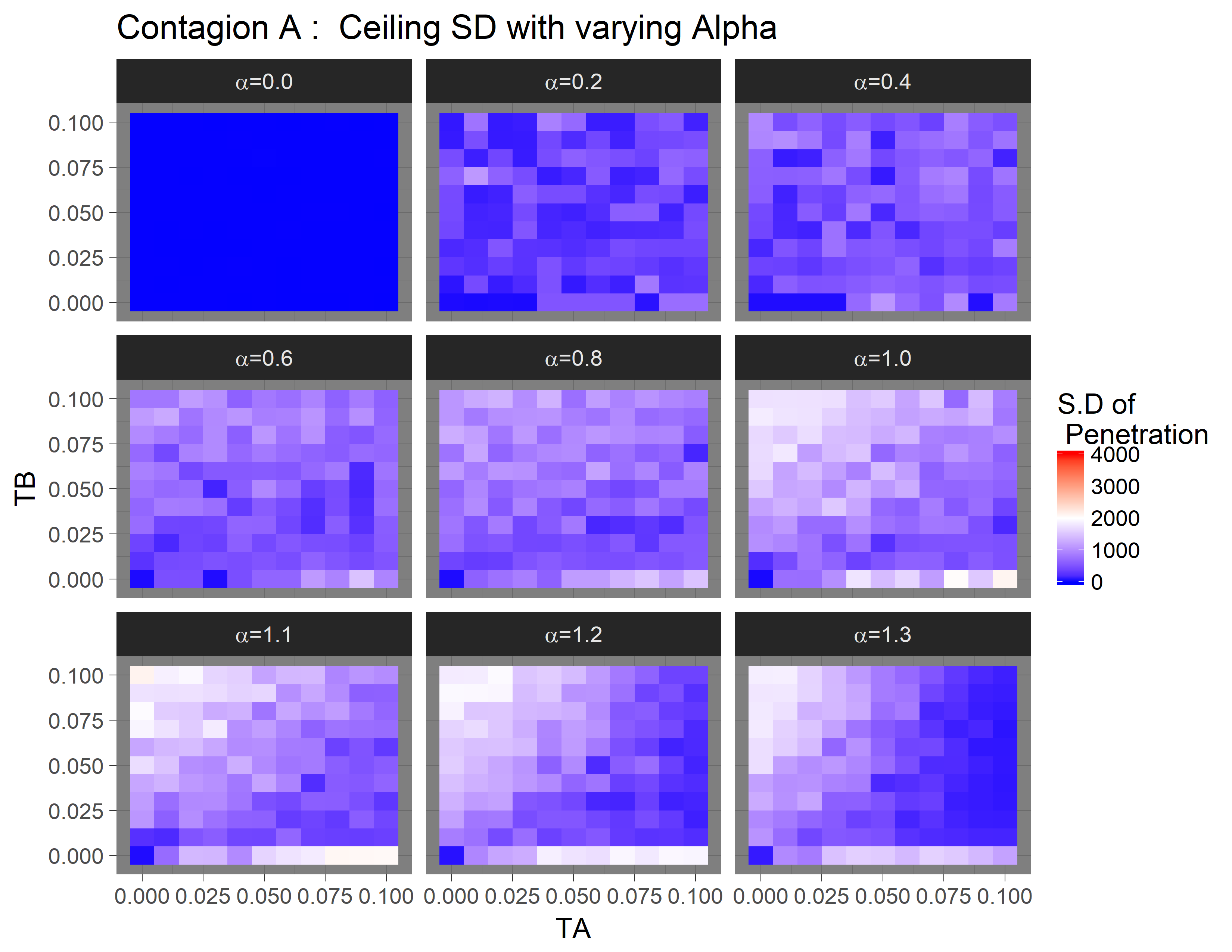}
        \caption{Multiplex Ceiling Standard Deviation of Contagion $A$.}{Since $A$ diffuses on the lattice, greatest instances of branching occur when the difference between $\TA$ and $\TB$ is large.}
        \label{fig:multiplex-A-C-std}
\end{figure}
 
Notice, the white regions on the upper left and bottom right corner. Further testing on single layer lattices (shown in the Figure~\ref{fig:Lattice-Diffusion-Curves} and Figure~\ref{fig:Lattice-KDE} in the Supplementary Materials) confirms the greatest branching occurs when the difference between $\TA$ and $\TB$ is large, which can branch into three clusters. Thus, the position of the white regions in Figure~\ref{fig:multiplex-A-C-std} confirm high levels of standard deviation as result of the branching. The strongest nonlinearities occur once more when $\TA >> \TB$ or $\TB >> \TA$, shown in Figure~\ref{fig:multiplex-A-C-std}. However, $\TB$ influences the diffusion variance of $A$ for many low values of $\TA$, shown by the top-left white region when $\alpha > 1.0$. In comparison, the effect of $\TA$ is most pronounced when $\TB = 0$. In other words, the dormancy variable of Contagion $B$ influences the ceiling of Contagion $A$ greatly because of its diffusion primacy. Also, note the bottom right corner has standard deviation always at 0, as it diffuses maximally.

\begin{figure}[!htb]
        \centering
        \includegraphics[width=1.0\linewidth]{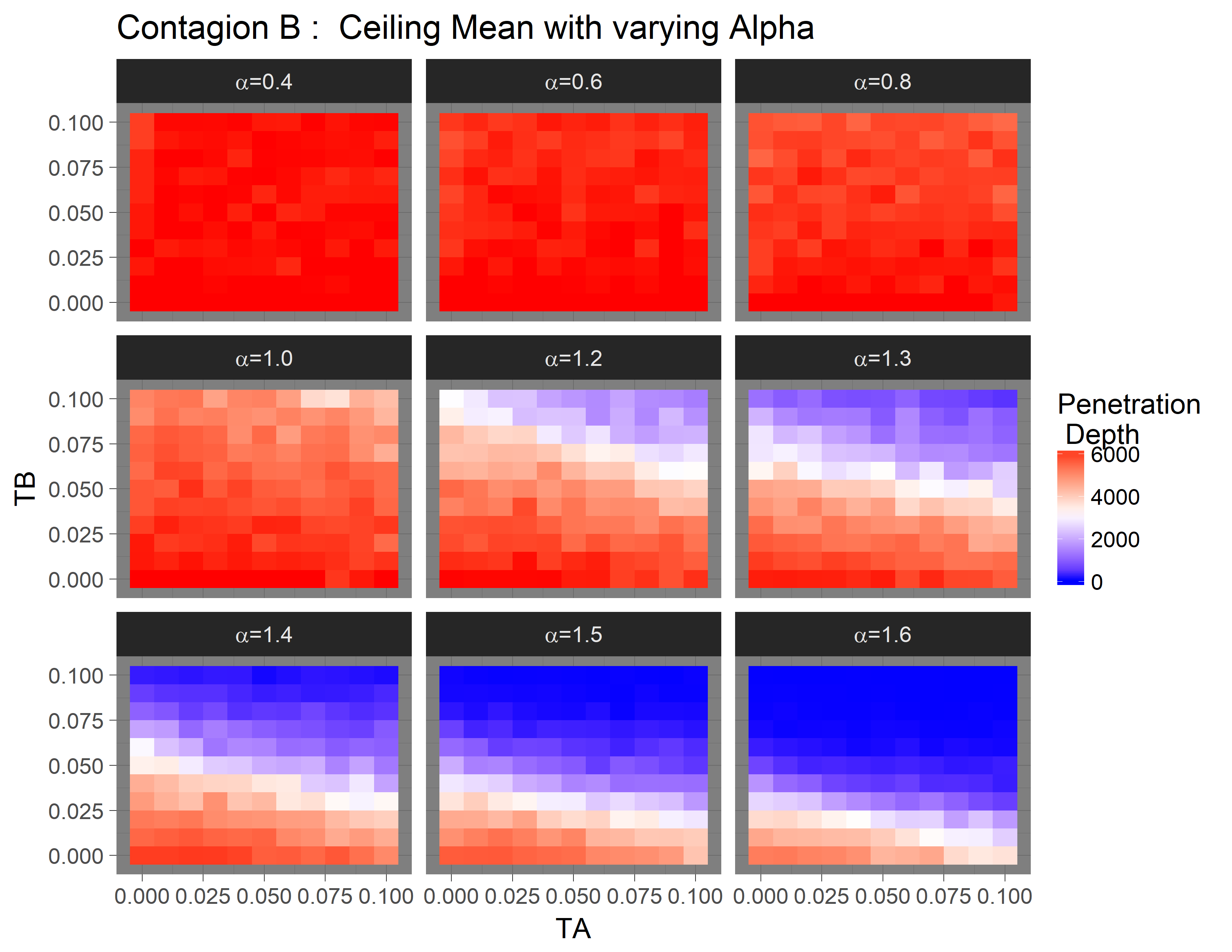}
        \caption{Multiplex Ceiling Mean of Contagion B}{The white line indicates a linear relationship for $\TA$ and $\TB$ to Contagion $B$'s ceiling sensitivity}
        \label{fig:multiplex-B-C-mean}
\end{figure}

Contagion $B$ produces a very different looking heat map in Figure~\ref{fig:multiplex-B-C-mean}. Given its fast diffusion, Contagion $B$ penetrates fully or close to fully for $\alpha < 0.8$. Then, a diagonal white line moves from the top right downwards and compresses towards the bottom left. The slope of this white line and its spread can be interpreted as Contagion $B$'s sensitivity to $\TA$. In other words, it quantifies how much the dormancy constant of Contagion $A$ affects its own diffusion. In a similar vein, the prior heat maps are also sensitivity analyses and also show the dominance of $\TB$. The effects of $\TA$ only come in play with sufficiently high $\alpha$ to slow the diffusion rate.

\begin{figure}[!htb]
        \centering
        \includegraphics[width=1.0\linewidth]{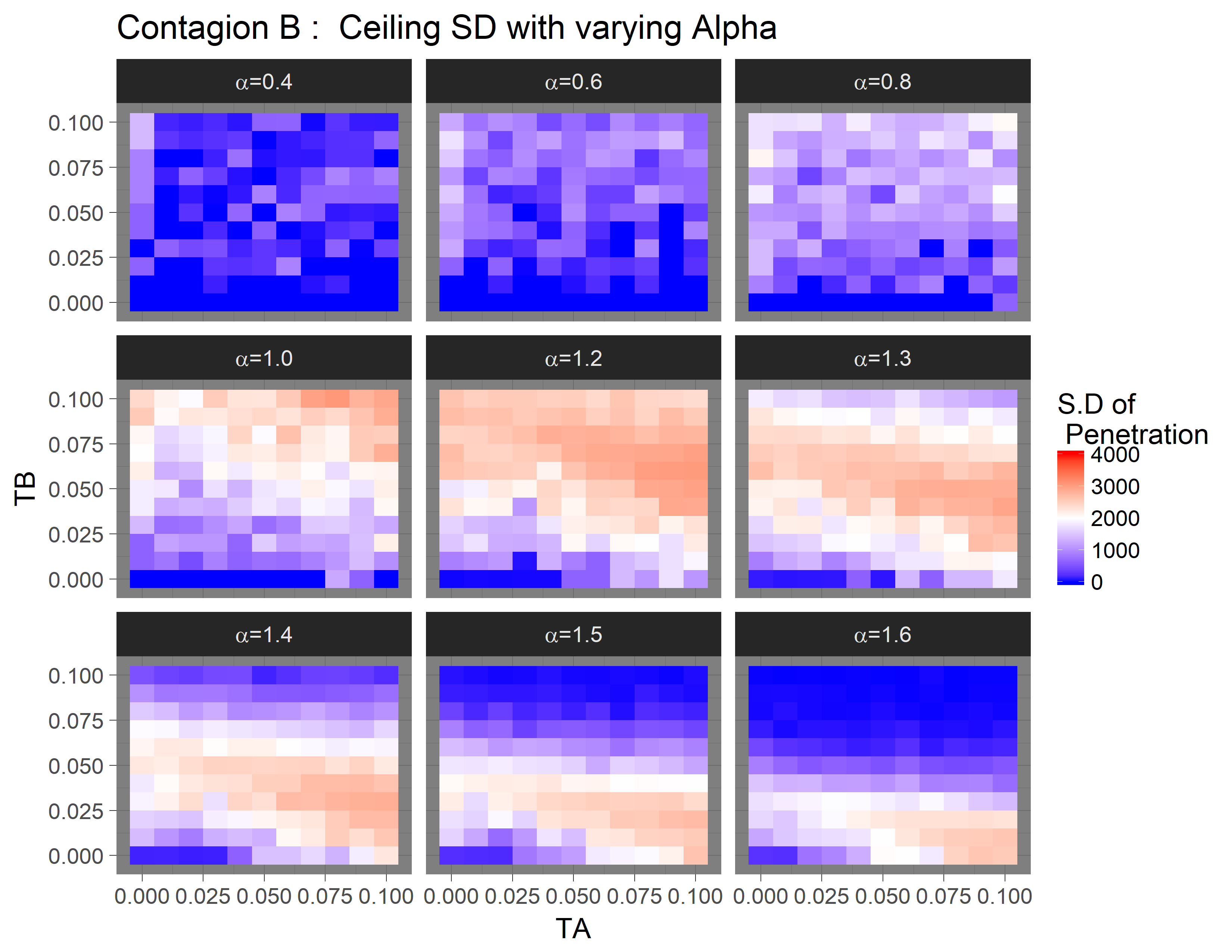}
        \caption{Multiplex Ceiling Standard Deviation of Contagion B}{The standard deviation for Contagion $B$ is greatest along the line in Figure~\ref{fig:multiplex-B-C-std}. The branching is linearly dependent on both $\TA$ and $\TB$.}
        \label{fig:multiplex-B-C-std}
\end{figure}

Having established that branching occurs we are interested in pinpointing the specific conditions that induce such instabilities. Instabilities can be inferred from high levels of standard deviation and we consider instability of Contagion $B$ shown in Figure~\ref{fig:multiplex-B-C-std}. The region of highest instability shown in red overlaps with the white line shown in Figure~\ref{fig:multiplex-B-C-mean}, the ceiling mean. This is particularly evident in the last row where $\alpha= 1.4$ to $1.6$. The region of the line becomes compressed and is sandwiched between two blue areas. The blue zone in the bottom left denotes the cases where Contagion $B$ fully diffuses, the blue zone above denotes the case where Contagion $B$ does not diffuse at all due to percolation produced by the high value of $\TB$. Increasing $\alpha$ produces the compression effect towards $\TA = \TB = 0$ from above as timescale increases.

One implication of the Contagion $B$ heat-map for standard deviation is that, unlike the lattice diffusion experiment, instabilities for Contagion $B$ not only occur when $\TA >> \TB$ or $\TB >> \TA$, but on any point of the white line. For instance, with each parameter set denoted with the tuple $(\alpha, ~\TA, ~\TB )$, the outcome of $(1.6,~0.00, ~0.04 )$ is equal to $(1.6, ~0.10, ~0.02 )$, although there is greater instability in the latter case when $\TA = 0.10$. Once more, we observe the "compression" effect towards $\TA=\TB=0$ from increasing $\alpha$. 

\begin{figure}[!htb]
        \centering
        \includegraphics[width=1.0\linewidth]{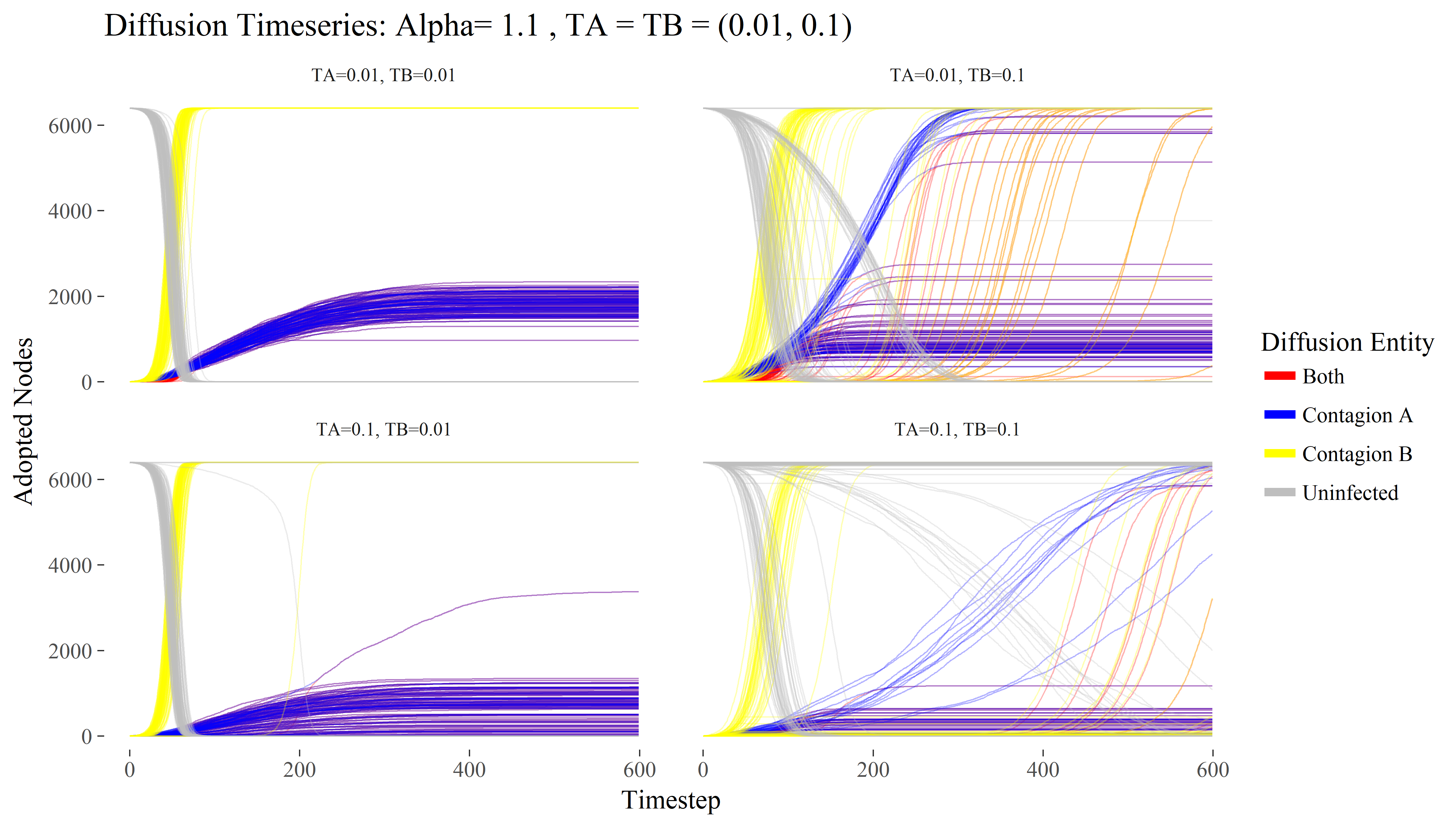}
        \caption{Interaction of $\TA$ and $\TB$}{}
        \label{fig:multiplex-diffusion-interaction}
\end{figure}

Lastly, we consider when both values of $\tau$ are non-negative and analyze their interaction. Figure~\ref{fig:multiplex-diffusion-interaction} demonstrates the interaction of $\TA$ and $\TB$ where we set low values equal to $0.01$ and high values at $0.1$. When both $\TA$ and $\TB$ are low, both converge to one cluster of ceilings. When $\TA$ is low and $\TB$ is high, the splitting effect is most pronounced. In other words, the number of Contagion $A$ curves that split upward is equal to the number of Contagion $B$ that do not diffuse at all or diffuse afterwards. This implies Contagion $A$ can produce synergistic effects that help boost the diffusion of $B$. For the most part, even when Contagion $B$ diffuses slower than Contagion $A$, it does not partial diffuse when $\alpha$ is sufficiently low, since it is not affected by local percolation. So even if the spread of Contagion $B$ is stymied within the RRG, once Contagion $A$ diffuses fully and synergistic effects take hold, it may still diffuse toward the maximal value defined by Contagion $A$ as shown in Figure~\ref{fig:multiplex-diffusion-interaction}b.

\section{Discussion} \label{sec:conclusion}
\subsection{Summary of Results}
The objective of this study is to investigate the properties of a newly proposed diffusion model by simulating two social contagions interacting on a multiplex network. We have determined how the parameters broadly influence the level of diffusion. Increasing $\alpha$ decreases the synergy as with the rate of diffusion. The stochastic dormancy constant of Contagion B lowers the penetration depth of both Contagion $B$ and Contagion $A$. We hypothesized that the parameter set $(\alpha, ~ \TA, ~\TB)$ is highly nonlinear. Diffusion of Contagion $A$ on the lattice branches into two or three clusters based on the ratio $\tau_A$ to $\tau_B$ for a set range of $\alpha$. Generally, when the difference between $\tau_A$ and $\tau_B$ is large within the region of $0.8<\alpha<1.3$, a branching is expected. 

Contagion $B$ diffuses much faster than Contagion $A$. It's faster diffusion is due to two reasons. First, the random-regular-graph's larger diffusion front compared to the lattice's, which is constrained physically. Second, it's diffusion is not constrained by local percolation. We conclude that primacy is important in this model of co-diffusion. Furthermore, if Contagion $B$ faces percolation from its own dormancy on the RRG, then Contagion $A$ can attain maximal penetration. If it diffuses more slowly than Contagion $B$, then it is subject to partial diffusion due to ring vaccination and the dormancy introduced by the faster diffusion contagion. This implies dormancy, this looser and one directional definition of immunity, is a sufficient condition for ring vaccination.

\subsection{Applications}
As we mentioned in the beginning, one critique of agent-based modeling is that its reductivism diminishes its utility and applicability. The interpretation in biology is clear: two diseases help each other spread by synergistically weakening the immune system of individuals, but once a certain number of hosts recover, the second disease has a harder time penetrating the populace due to diminished density. However, there are applications outside of biology. Here, we offer some potential ways to interpret these results, in particular, the notions of synergy and dormancy, with GPTs mentioned in the literature review in Section~\ref{sec:lit-review}.

Blockchain technology has recently been considered as a potential general purpose technology~\cite{catalini2016some}~\cite{kane2017blockchain}~\cite{davidson2016economics}. The application of blockchain to cryptocurrencies, or Bitcoin, spurned a large wave of interest and investment in 2017~\cite{carrick2016bitcoin}~\cite{bitVol}. 

Given the speculation surrounding cryptocurrency, this investment behavior can certainly be categorized as a social contagion~\cite{guardian_2017}. While cryptocurrency theoretically serves as means substance for transactions, current investors treat it more as a commodity or asset, rather than as liquid money. In that regard, the adoption of cryptocurrency is not dissimilar to a firm's adoption of a general-purpose technology to increase output. In essence, there is an investment in the GPT, then an expected benefit or return from the adoption. If you are an investor with a finite portfolio, there are two questions that naturally arise. First, do you adopt the technology and invest in Bitcoin? Second, which cryptocurrency do you choose to invest in? 

The first question corresponds to Equation~\ref{eq:naive-adoption}, where all cryptocurrencies synergistically contribute to the diffusion of the adoption behavior. The second question refers to the coin-flip in Equation~\ref{eq:A-B-coinflip}, which frames individual cryptocurrencies as competitors. However, this does not preclude the possibility of adopting the one not chosen at a future time. In this regard, the diffusion mechanism of this paper captures both cooperative and quasi-competitive aspects of the technology by Chandrasekaran~\cite{chandrasekaran2007critical}.

\subsection{Future work}
There are multiple pathways for future inquiry that build on the weaknesses in this study, specifically with shape parametrization, seeding, the diffusion mechanism, and graph topologies. In this study we assume the shape characterization of contagion adoption to be equal; that is, in our parametrization $K_A=K_B$. Changing this parameter would allow for a more in-depth study of the effects of primacy by controlling precisely how much faster one of the contagions diffuses relatively to the other. Controlling for $K_B$ would be meaningful for interdependent networks, such as controlling for the much faster diffusion on the random-regular-graph. For $\TA$ to restrain the diffusion of $\TB$, $B$ must diffuse slowly. Due to the difference in topology, the current parameter pairs are insufficient to consider the case where Contagion $B$ diffuses slower than $A$. The timing of entry is an interesting question. Given that Contagion $A$ diffuses first with a high value of $\tau_A$, a late entry by Contagion $B$ would most likely affect its penetration depth. Quantifying this result would be useful for benchmarking the benefits or risks of late adaptation of general purpose technologies.

Different topologies would certainly yield different results. We have shown a difference in the diffusion of spatial and long-range graphs. For a model to accurately describe how social media and physical newspapers help each other spread news, more precise networks would have to be implemented~\cite{gallos2012people} to capture the mode or dynamical channels. In this case,  power-law graphs and lattice graph may be suitable. Methods for extracting the ceiling such as unsupervised learning to produce clusters would help understand the modality of convergent behavior.

Another avenue for research is modeling exclusive adoption as we have outlined in Section~\ref{sec:excl-adoption}. Such a probability kernel for diffusion would be useful for modeling products where mutual adoption is impossible. There are also many limitations to threshold models. Instead, a network coordination model could be used to produce a more complicated, but still inherently rational, diffusion mechanism. 

\section*{Acknowledgement}
We'd like to thank the Discovery Cluster at Dartmouth College, the Neukom Institute and the Jack Byrne Scholarship Fund for funding, and the Department of Quantitative Social Sciences for feedback. We'd like to thank David Cottrell and Jennifer Kuo for statistical consulting, and Sean Westwood for suggestions.

\newpage

\section*{Supplementary Materials}
\subsection{Clustering of Diffusion Ceilings}
\begin{figure}[!htb]
        \centering
        \includegraphics[width=0.5\linewidth]{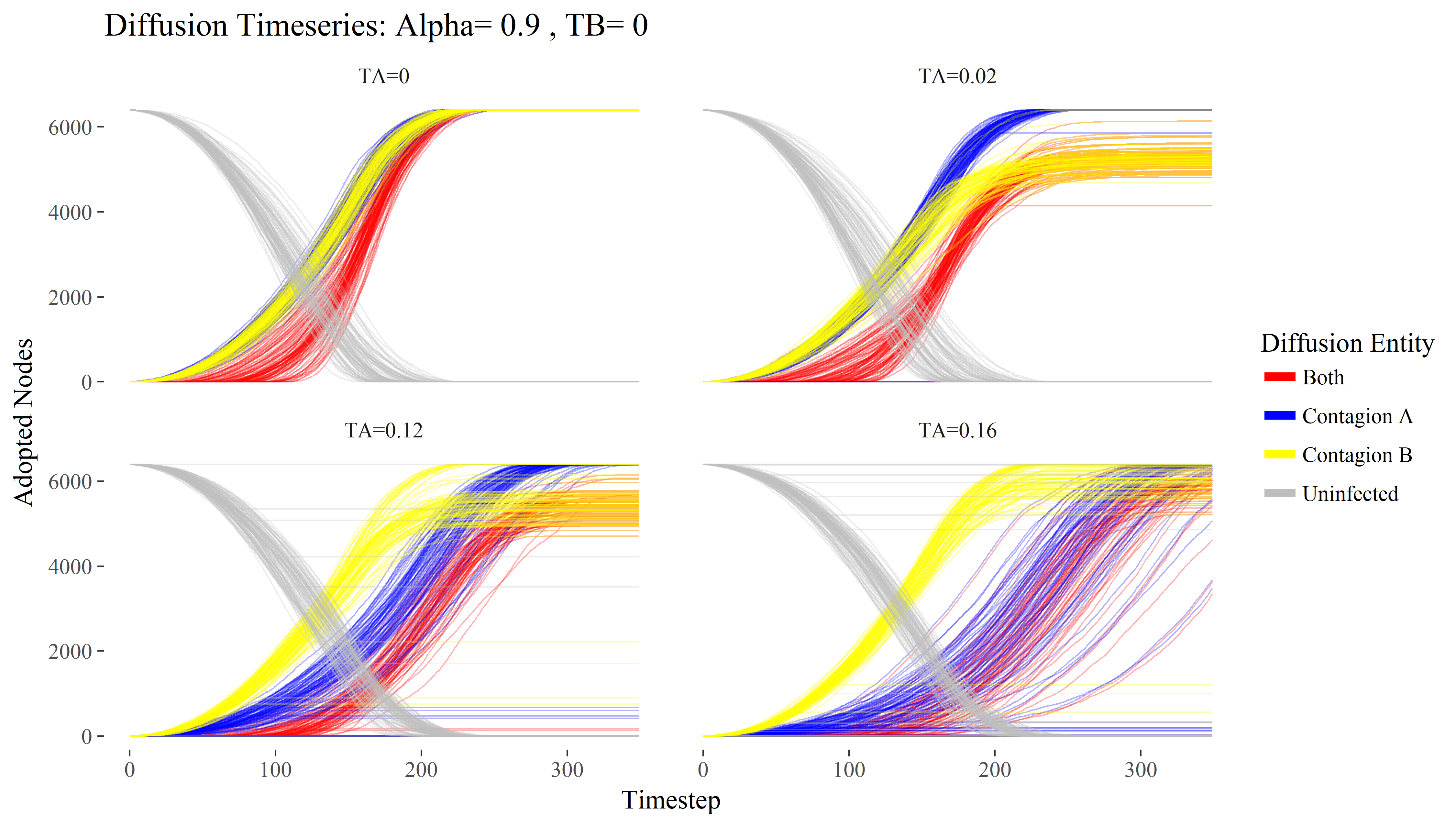}
        \caption{Diffusion Curves while Varying $\tau_A$}{The value of $\TA$ increases from $0$ to $0.16$. When $\tau_A = 0.02$ the rate of diffusion curve of $A$ does not decrease by much. This is evident as the blue curve does not shift to the right appreciably. However, nodes infected by $A$ no longer participate in the diffusion process and thus the penetration depth of $B$ is diminished. When $\tau_A$ is high, the rate of diffusion of Contagion $A$ slows down sufficiently such that Contagion $B$ diffuses more quickly and can thus fully diffuse.}
        \label{fig:Lattice-Diffusion-Curves}
\end{figure}

\begin{figure}[!htb]
        \centering
        \includegraphics[width=0.5\linewidth]{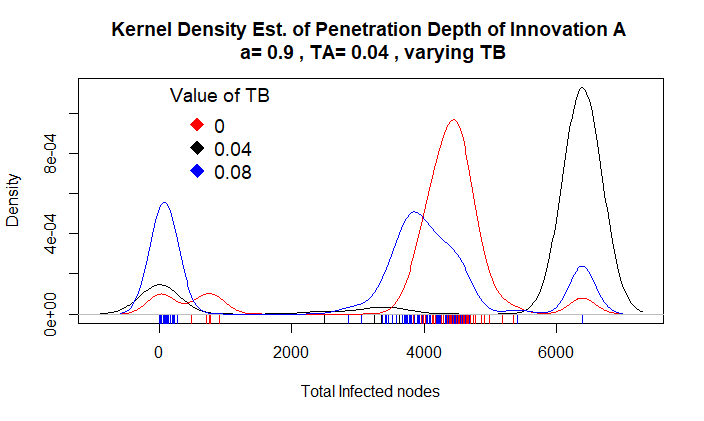}
        \caption{Kernel Density of Lattice Ceilings}{A difference of $0.04$ between $\TA$ and $\TB$ produces a tri-modal distribution of ceilings, while when they are equal the distribution is bi-modal.}
        \label{fig:Lattice-KDE}
\end{figure}

This instability can produce more than two clusters. As shown in the kernel density estimate of Figure~\ref{fig:Lattice-KDE}, a trimodal distribution is possible, where we vary $\tau_B$ relative to $\tau_A$. When $\tau_A = \tau_B = 0.04$ the distribution remains bimodal, and the graph shows that Contagion $A$ diffuses either completely or almost not at all. 

We then confirm this for the multiplex case with the following evolution of kernel density estimates.

\begin{figure}[!htb]
        \centering
        \includegraphics[width=1.0\linewidth]{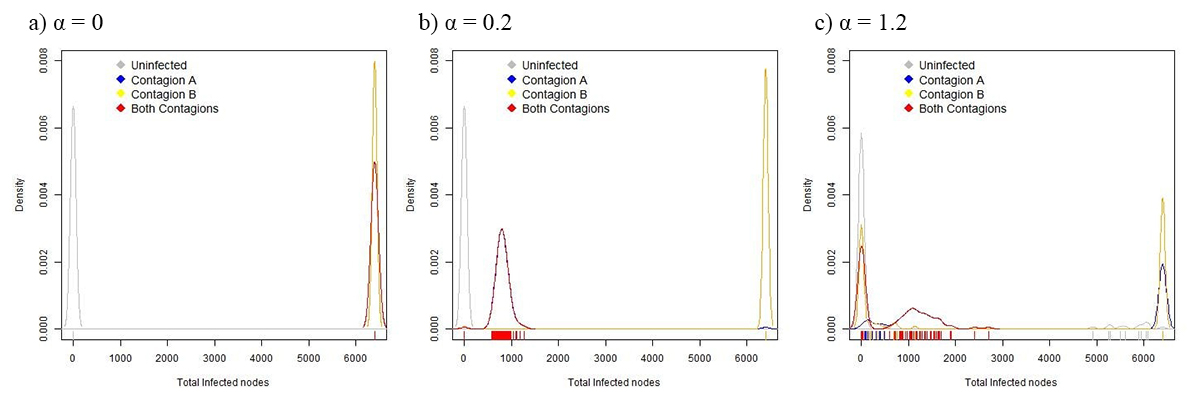}
        \caption{KDEs of Diffusion Ceilings while fixing $\TA=0$ and $\TB=0.08$. Diffusion of Contagion $A$ can be unimodal or bimodal.}
        \label{fig:Sup-KDE-alpha}
\end{figure}

\begin{figure}[!htb]
        \centering
        \includegraphics[width=1.0\linewidth]{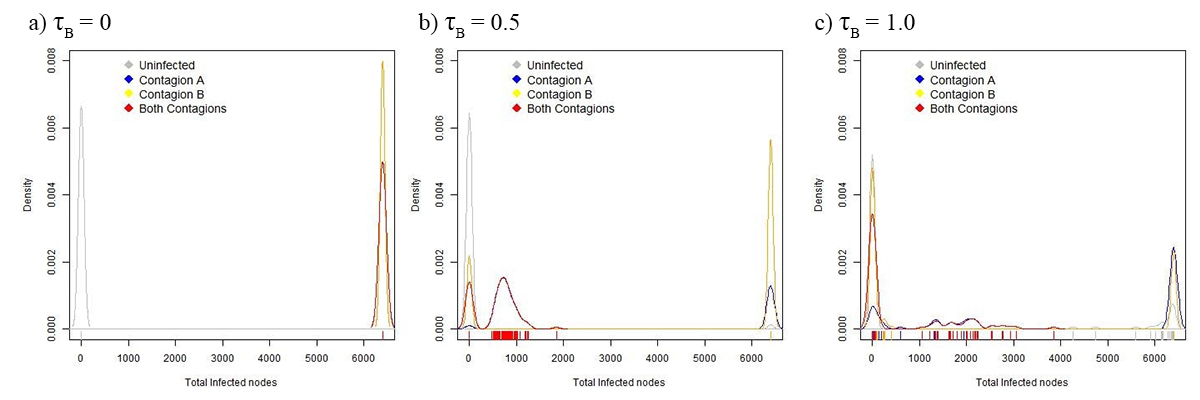}
        \caption{KDEs of Diffusion Ceilings while fixing $\alpha=1.3$ and $\TA=0$. Diffusion of Contagion $A$ can be unimodal, bimodal or trimodal.}
        \label{fig:Sup-KDE-TB}
\end{figure}

\subsection{Theoretical Derivation in Well-Mixed Populations}
We derive an expression for the rate of co-diffusion in a well-mixed population with degree $\kappa$. We define the adoption rate of Contagion $j$ as $\phi_j$ and we denote the adoption of $A$ and $B$ as ${AB}$. Thus explicitly:
$$
\begin{cases}
\phi_A 		& \defeq \textit{Adoption Rate of A} \\
\phi_B 		& \defeq \textit{Adoption Rate of B} \\
\phi_{AB} 	& \defeq \textit{Adoption Rate of A and B}
\end{cases}
$$
Next, define the proportion of each type as $x_A$, $x_B$, $x_{AB}$, $x_\emptyset$, and $x_R$, where the subscript $\emptyset$ denotes naive individuals and $R$ denotes dormant individuals. Note that $x_A$, $x_B$ and $x_{AB}$ only include naive individuals that--- nodes that are dormant. Now we state our mean-field equations:
\begin{equation} \label{eq:mean-square}
\begin{cases}
1 				&= x_A + x_B + x_{AB} + x_\emptyset + x_R \\
\dot{x}_A 		&= x_\emptyset \phi_A(x_A,x_B) - \tau_A x_A  \\
\dot{x}_B		&= x_\emptyset \phi_B(x_A,x_B) - \tau_B x_B \\
\dot{x}_{AB} 	&= x_A \phi_{AB}(x_A,x_B) +x_B \phi_{AB}(x_A,x_B) - \tau_{AB} x_{AB} \\
\dot{x}_R		&= \TA x_A + \TB x_B + \tau_{AB} x_{AB}
\end{cases}
\end{equation}

In layman language, the first equation states that there are five groups whose proportions sum to one. Equations two through five denote the rate of change for each respective sub-group. The rate of change for $A$, denoted $\dot{x}_A$ is equal to the adoption rate of $A$ multiplied by the total proportion of $A$, subtracted by $\tau_A x_A$, the total number of nodes that go dormant. It is the same case for Contagion $B$ and both. The last equation describes the nodes that go dormant, which is the sum of all the $\tau$'s multiplied by their respective sub-groups.

\subsection*{References}
\bibliographystyle{vancouver}
\bibliography{refs}

\end{document}